# Enhancing Collaboration for Software Engineers through Matching

**Nayaab Azim**
Computer Science
Virginia Tech
Blacksburg VA USA
nayaab@vt.edu

**Sadath Ullah Khan Mohammed**
Computer Science
Virginia Tech
Blacksburg VA USA
msadath@vt.edu

**Evan Phaup**
Computer Science
Virginia Tech
Blacksburg VA USA
phaupe22@vt.edu

**Adeyemi Aina**
Computer Science
Virginia Tech
Blacksburg VA USA
ainababs0@vt.edu

## ABSTRACT

In recent years, the field of software engineering has experienced a considerable increase in demand for competent experts, resulting in an increased demand for platforms that connect software engineers and facilitate collaboration. In response to this necessity, in this paper we present a project to solve the lack of a proper one-stop connection platform for software engineers and promoting collaborative learning and upskilling. The idea of the project is to develop a web-based application (NEXAS) that would facilitate connecting and collaborating between software engineers. The application would perform algorithmic matching to suggest user connections based on their technical profiles and interests. The users can filter profiles, discover open projects, and form collaboration groups. Using this application will enable users to connect with peers having similar interests, thereby creating a community network tailored exclusively for software engineers.

## KEYWORDS

Matching algorithms, collaborative software engineering, user preferences,

## 1. INTRODUCTION

With rapid changes in software technology in the 21st century, Software Engineers have the task of constantly upgrading their skillset. Learning new technologies can be daunting however groups and community are a great offset. Recent studies have shown multiple benefits of collaborative learning among software engineers such as organized, time-effective learning and guidance among others (Layman, 2006). Although Software Engineers meet their peers on a regular basis, most of them are unaware of each other's skill sets. Thus, when an engineer experiences difficulty in a project or a particular technology niche, he is left wondering.

The proposed solution NEXAS web-application mitigates the above-mentioned challenges, this is a connection platform specifically designed for software developers. The users would be able to connect with other software engineers in the community. Once the user sets up his/her profile and technology interests, the application would then suggest connections by using matching algorithms on parameters such as project interests, proximity, learning outcome and experience levels. Moreover, the users can also find and view technical profiles of other users by applying desired filters. Discord, Stack overflow, and other platforms exist to facilitate collaboration among software engineers. These platforms, however, lack the ability to match users based on proximity and skill set, which NEXAS aims to provide.

### 1.1 Matching Algorithm

Matching algorithm are a critical component of the proposed Software Engineers connection platform; based on their technical profiles, project interests, and other criteria such as experience levels and learning outcomes, these algorithms will allow the program to identify suitable connections for Software Engineers. The use of matching algorithms will ease connection for Software Engineers and increase the chances of successful collaboration by suggesting relevant and compatible connections. Therefore, the design and implementation of effective matching algorithms will be a key focus.

## 2. RELATED WORK

Matching social applications have gained interest in software development collaboration, and several related studies have explored various aspects of this topic. Liu et al. (2009) provided an overview of algorithms for rule generation and matchmaking in context-aware systems. The paper discusses various techniques for matching users with relevant content or services based on contextual information, using rule-based and machine learning algorithms. The authors also explored techniques for combining these two approaches to create more efficient systems with better accuracy rates.

Wang and Li (n.d.) proposed an approach based on Hebbian Learning Law that can automatically group distributed e-learners with similar interests, enhancing collaborative learning among them. This algorithm avoids the need for difficult design work required for user preference representation or user similarity calculation, making it suitable for open and distributed environments like e-learning platforms. Experimental results show that this algorithm has preferable prediction accuracy as well as improved scalability and satisfaction levels compared to existing approaches in the field of e-learning social network exploiting approach.

Other related studies have examined specific features or use cases of proximity-based social applications for software development collaboration. For example, Ali et al. (2017) proposed a dynamic self-organizing community or group mechanism with an improved matchmaking algorithm. The system can group similar agents according to their preferences and capabilities, and the authors discuss how to represent, evaluate, exchange, and matchmake during the self-organizing process. A prototype was built for verification purposes, and experiments based on real learner data show that this mechanism can organize learners properly while improving speed and efficiency of searches for peer students owning relevant knowledge resources.

## 3. RESEARCH GOALS AND OBJECTIVES

This application would enhance collaborative learning and help build a community of software engineers. It is a 2-way solution: people looking to join a project can find it, on the other hand open projects looking to add new members could do so too. Users could also discuss doubts, logic and process flow by utilizing the text messaging feature. This would result in time efficient work rather than a single developer being stuck at an issue. Furthermore, it would develop a sense of team-work spirit among the users, which is one of the key attributes to become a great software engineer [2](Li et al., 2020).

## 4. ARCHITECTURAL DESIGN

The design architecture of our site using the MERN stack typically involves a client-server architecture, with the client being built using React and the server being built using Node.js and Express. Here is a high-level overview of the different components and layers of the architecture:

1. **Client Layer** - This layer is responsible for rendering the user interface and handling user interactions. It is built using React, which allows for the creation of reusable components and efficient rendering of UI elements.
2. **API Layer** - This layer serves as an interface between the client layer and the server layer, providing a set of RESTful endpoints for the client to interact with. The AP layer is built using Express and is responsible for handling incoming requests from the client and sending back appropriate responses.
3. **Server Layer -** This layer is responsible for processing incoming requests from the API layer and interacting with the database layer to retrieve or store data. The server layer is built using Node.js and Express, and it will incorporate middleware for handling authentication, logging, error handling and other features.
4. **Database layer -** This layer is responsible for storing and managing the site data. The database will include tables for storing user profiles, chat messages, and other user-generated data. The database layer is built using MongoDB, which is a popular NoSQL database that works well with Node.js and Express.
5. **Hosting -** This layer is responsible for hosting the site on the service like Amazon Web Services (AWS) and manages the various components of the site.
6. **State Management -** To manage the state of the client layer and ensure that data is efficiently passed between different components and updates are handled in a consistent manner, we will use a tool called Redux.
7. **Security -** Implementing third-party authentication service called Firebase to handle authentication and prevent unauthorized access or malicious attacks
8. **Message Queue -** Using a message queue or real-time communication tool like socket.io to enable real-time messaging and notifications between users.
9. **Location API -** Using a service like Google Maps API or OpenStreetMap to add location-based search functionality.
10. **Testing -** Using testing framework like Jest to test the site and ensure that it's working as expected.
11. **Logging -** Using logging and monitoring tools to track site performance, identify issues, and make improvements over time.

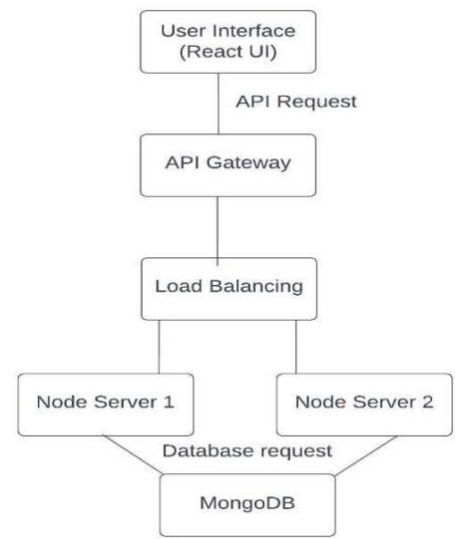

*Figure 1: High-Level Architecture Flow Diagram*

**Enhancing Collaboration for Software Engineers through Matchmaking and Proximity**

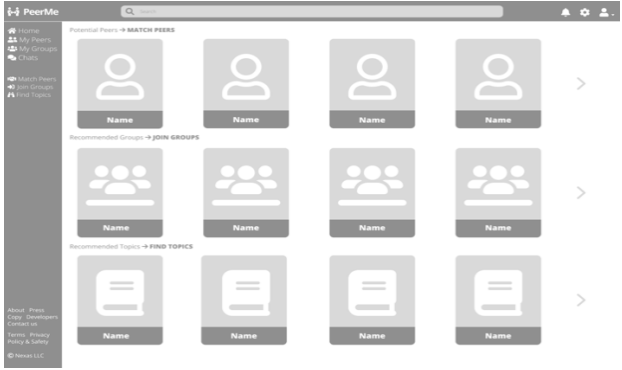

**Figure 1: user's landing page with preferences to match peers, join groups find personalized topics**

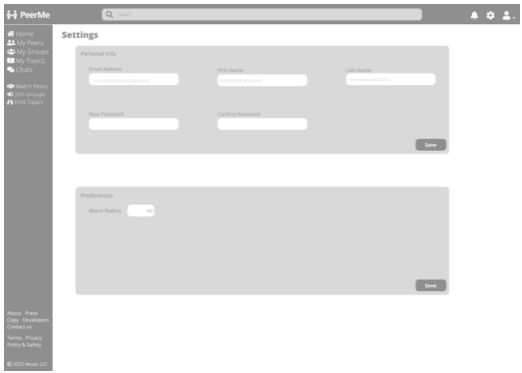

**Figure 2: user's personalized settings menu, showing matching radius for proximity preference.**

## 5. IMPLEMENTATION PROCESS

In the implementation process, we carried out the following steps to successfully develop and deploy the Nexas application for engineers:

**Requirement Gathering:** We analyzed the project's requirements, focusing on the functionality, performance, and benefits of the matching system for engineers. We used GitHub issue tracking to manage these requirements and facilitate the project's activity.

**Reviewed related projects:** We conducted an extensive review of related projects and existing solutions within this domain. This step allowed us to identify best practices and challenges encountered in related applications, which helped us incorporate valuable insights into both our architectural design and development approach. We established that the application had a sound foundation and employed proven techniques for streamlining the process of pairing engineers.

**Development:** For the application, the technology stack utilized was Node.js for backend development, MongoDB for data storage, and React.js for frontend development. We utilized the initial architectural base as discussed at the proposal stage, also Git to manage version control of Nexas code base, creating separate branches for different features and merging them after thorough testing and review.

The machine learning algorithm implemented for matching engineers was the K-nearest neighbor (KNN) algorithm, which facilitated accurate and efficient matching of software engineers based on their skills and expertise.

**Testing:** We conducted unit, integration, and system testing to ensure the application's functionality and performance met the project requirements. Testing frameworks and tools we have included Jest for automated testing in the current code base.

## 6. TESTING MODULE

We employed a comprehensive testing process to ensure the quality and reliability of the Nexas application.

**Unit Testing:** We tested individual functions and modules to ensure the intended functionality was expected and met the project requirements. Tools like Jest were used to facilitate this process. We have included several test cases and scenarios in the code, and test controller for unit testing.

**Integration Testing:** We tested the integration of different modules and components, controllers to the backend to the database to ensure they worked together seamlessly and as expected. Postman was mostly used as the tool for API testing during this phase.

**System Testing:** The application was tested to ensure it met the project requirements and functioned as expected.

**Security Testing:** We tested the application's security measures to ensure they protected user data and prevented unauthorized access also, with Netlify's full security suite.

## 7. PROCESSES USED:

In our software project, we utilized an agile methodology to manage the development process. We divided the project into several sprints, with each sprint focusing on a specific set of features or tasks. At the beginning of each sprint, we held a sprint planning meeting where we discussed the objectives and goals for the upcoming sprint. During the sprint, we held daily stand-up meetings to keep everyone on the team informed about the progress and any roadblocks that needed to be addressed. At the end of each sprint, we held a sprint review meeting where we demonstrated the completed work to stakeholders and obtained their feedback. This iterative and collaborative approach allowed us to quickly adapt to changing requirements and feedback, resulting in a high-quality and effective software product.

# 8 DEPLOYMENT AND MAINTENANCE PLAN

## 8.1 Deployment Plan

Our Match-making website for Software developers will use a rolling deployment strategy to ensure high availability and minimal downtime during the deployment process. This strategy can help reduce downtime and minimize the risk of application failures when there will be huge incoming users by ensuring that only a small number of servers are affected at a time.

Our deployment process starts with local testing using the Rest APIs. We tested and debugged our code in a local environment to ensure that it meets our quality standards and is free from errors. Once our code has passed local testing, it will be deployed to a staging environment, where further testing will be conducted to identify any issues that may arise.

Once the staging environment has been thoroughly tested and is deemed stable, we will use Netlify to deploy the updated version to production. Netlify is a cloud-based platform that simplifies the deployment process by providing automated builds, continuous deployment, and serverless functions. Netlify is a cost-effective and efficient way to deploy our application to production, as it provides high availability, scalability, and security. We structured our git-hub repo in two different branches which are back-end and front-end to have error-free deployment.

During the deployment process, we will use a ROLLING DEPLOYMENT strategy to ensure that the website remains accessible to users. The updated version will be deployed gradually, one server at a time, while the existing version continues to handle user traffic. This approach will help to reduce the risk of application failures and minimize downtime during the deployment process.

## 8.2 Maintenance Process:

**Regular Updates:** To ensure that the website is up-to-date and functioning correctly, we will regularly update. This includes updating the server software, security patches, and bug fixes.

**Monitoring:** We will monitor the website to ensure that it is functioning correctly and efficiently. This includes monitoring the server load, user traffic, and database performance.

**Security:** To ensure that the website is secure, we will conduct regular security audits. This includes checking for vulnerabilities, updating passwords, and implementing security protocols.

**User Feedback:** User feedback will be regularly collected to ensure that the website is meeting the needs of its users. This feedback will be used to improve the website's features, functionality, and user experience.

**Backup and Disaster Recovery:** Regular backups will be conducted to ensure that the website's data is secure and can be recovered in the event of a disaster. This includes implementing disaster recovery protocols and data backup procedures.

# 9. LIMITATIONS:

While our application provides valuable steps toward resolving the problem at hand, some limitations to our study should be noted. Firstly, the match-making algorithm is trained on a limited number of features and is prone to overfitting. Furthermore, the features used for training have all been assigned equal weights, which is not the case in real-world scenarios as some features such as domain and programming languages may be more relevant than others such as years of experience. Finally, due to time constraints, we were unable to get users to try all the features and functionality of the software, which may limit the scope of our findings. Despite these limitations, our research provides a baseline platform for collaboration through matchmaking among software engineers and highlights areas for future research and improvement.

# 10. FUTURE WORK

## 10.1 Proximity preference

Proximity preference is an innovative approach that connects users with other software engineers in proximity. This allows users to form collaborative groups with other professionals in their area, making in-person meetups and skill-sharing sessions easier to organize. The NEXAS web-application can use proximity feature to encourage software engineers to take advantage of their local network and strengthen connections with like-minded professionals within proximity.

## 10.2 Multi-platform Access

The Nexas project currently only focuses on a web-based application, hence making efforts in other platforms such as Android and iOS would be a good direction to proceed. Another avenue for future research could be to get the algorithm running on a larger and more diverse user base to better understand the software's usability across different user groups and contexts. Additionally, further investigations could focus on exploring the messaging features, and project group features, which would make the collaboration among software engineers even better. The project groups would enable users to view the open projects in the community and join the project team that intrigues them. Finally, our research has only been conducted on a local host, so deploying the application on a server will be required in the next steps to gain a more comprehensive understanding of its capabilities. Overall, there are many opportunities for future research to build on our findings and continue to improve the performance and usability of the software.

## 11. CONCLUSION

In conclusion, the increasing demand for competent software engineers has led to the need for a platform that connects them and fosters collaboration. The proposed project, NEXAS, aims to address this need by providing a one-stop platform for software engineers to connect, collaborate, learn, and upskill. With algorithmic matching and project discovery features, NEXAS has the potential to revolutionize the software engineering industry and enable engineers to work together more effectively.

The successful development and deployment of the Nexas application for engineers were made possible through a structured implementation process. Requirement gathering was conducted, and related projects were reviewed to ensure the application's functionality and benefits were aligned with industry best practices. Development involved using Node.js for the backend, MongoDB for data storage, and React.js for frontend development, with Git employed for version control. The K-nearest neighbor algorithm was implemented for efficient and accurate matching of software engineers. Finally, comprehensive unit, integration, and system testing were carried out using frameworks such as Jest to ensure optimal functionality and performance.

The NEXAS app offers several benefits for software engineers. Firstly, it provides a specialized platform designed specifically for software developers, enabling them to connect with others in their community. The application's algorithmic matching capabilities suggest connections based on project interests, proximity, learning outcomes, and experience levels, streamlining the process of finding potential collaborators. The ability to filter the technical profiles of other users further enhances the user experience. Unlike other collaboration platforms such as Discord and Stack Overflow, NEXAS offers the unique benefit of matching users based on skill set, facilitating more effective collaboration opportunities. Ultimately, the NEXAS app provides software engineers with a one-stop-shop platform for connection, collaboration, learning, and upskilling.